\documentclass[twoside,twocolumn,10pt]{article}

\usepackage{wscg}
\RequirePackage{ifpdf}
\ifpdf
 \RequirePackage[pdftex]{graphicx}
 \RequirePackage[pdftex]{color}
\else
 \RequirePackage[dvips,draft]{graphicx}
 \RequirePackage[dvips]{color}
\fi

\usepackage{nopageno}
\usepackage{url}
\usepackage{amsmath}
\usepackage[linesnumbered,ruled]{algorithm2e}

\SetKwRepeat{Do}{do}{while}

\title{Performance Assessment of Diffusive Load Balancing for Distributed Particle Advection}

\author{
\parbox{0.25\textwidth}{\centering
Ali Can Demiralp\\[1mm]
RWTH Aachen\\
Kopernikusstrasse 6\\
52074, Aachen, Germany\\[1mm]
}
\hspace{0.05\textwidth}
\parbox{0.35\textwidth}{\centering
Dirk Norbert Helmrich\\[1mm]
Forschungszentrum J\"ulich GmbH\\
Wilhelm-Johnen-Strasse\\
52428, J\"ulich, Germany\\[1mm]
}
\hspace{0.05\textwidth}
\parbox{0.25\textwidth}{\centering
Joachim Protze\\[1mm]
RWTH Aachen\\
Kopernikusstrasse 6\\
52074, Aachen, Germany\\[1mm]
}
\\\\
\hspace{0.05\textwidth}
\parbox{0.25\textwidth}{\centering
Torsten Wolfgang Kuhlen\\[1mm]
RWTH Aachen\\
Kopernikusstrasse 6\\
52074, Aachen, Germany\\[1mm]
}
\hspace{0.05\textwidth}
\parbox{0.25\textwidth}{\centering
Tim Gerrits\\[1mm]
RWTH Aachen\\
Kopernikusstrasse 6\\
52074, Aachen, Germany\\[1mm]
}
}

\makeatletter
\def\Uslash{\mathbin{\mathchar`\/}\@ifnextchar{/}{\kern-.15em}{}}
\g@addto@macro\UrlSpecials{\do \/ {\Uslash}}
\def\Ucolon{\mathbin{\mathchar`:}\@ifnextchar{/}{\kern-.1em}{}}
\g@addto@macro\UrlSpecials{\do : {\Ucolon}}
\makeatother

\begin{document}
\twocolumn[{\csname @twocolumnfalse\endcsname
\maketitle

\begin{abstract}
\noindent
Particle advection is the approach for extraction of integral curves from vector fields. Efficient parallelization of particle advection is a challenging task due to the problem of load imbalance, in which processes are assigned unequal workloads, causing some of them to idle as the others are performing compute. Various approaches to load balancing exist, yet they all involve trade-offs such as increased inter-process communication, or the need for central control structures. In this work, we present two local load balancing methods for particle advection based on the family of diffusive load balancing. Each process has access to the blocks of its neighboring processes, which enables dynamic sharing of the particles based on a metric defined by the workload of the neighborhood. The approaches are assessed in terms of strong and weak scaling as well as load imbalance. We show that the methods reduce the total run-time of advection and are promising with regard to scaling as they operate locally on isolated process neighborhoods. 
\end{abstract}


\subsection*{Keywords}
Particle Advection, Distributed Algorithms, Load Balancing
\vspace*{1.0\baselineskip}
}]

\section{Introduction}
\copyrightspace

Particle advection is an important method for analysis and visualization of vector fields. The idea is to place a set of (often massless) particles within the vector field, and integrate them over time using a numerical integration scheme such as the Runge-Kutta family of methods. Aside from its primary use for estimation of integral curves in vector data, it serves as a basis for various feature extraction methods such as Lagrangian Coherent Structures \cite{Haller2000}.

Modern vector datasets vary from several gigabytes to the upper end of the terabyte range. 
Considering the size of the data, and the number of particles necessary to represent the domain, parallel compute becomes essential and is in fact a standard tool for particle advection today \cite{Pugmire2009, Peterka2011}.

Parallel particle advection suffers from the problem of load imbalance, in which the vector field blocks and/or the particle sets are distributed in an unfair way, leading to some processes idling as others are performing compute. This is economically undesirable as the idle processes waste core-hours, yet are still paid for by the application user in the form of allocated cluster resources or currency. It is also desirable to have low advection times in order to enable real-time adjustments to the particle set and advection parameters. Due to these aspects, many recent approaches focus on the problem of load balancing \cite{Zhang2018a, Zhang2018b, Binyahib2019a, Binyahib2021}.

Although a variety of load balancing approaches already exist, they invariably come with trade-offs. The approach of \cite{Pugmire2009} offers decent performance, yet involves a task dispatching model that is challenging to implement in a distributed environment. The approach in \cite{Zhang2018a} involves transmission of the vector field across processes which may cause significant communication overhead. The approach presented in \cite{Zhang2018b} involves a parallel K-D tree construction at each round. The approach of \cite{Binyahib2019a} is shown to maximize resource usage, however may involve a set of serial communications until work is delivered to the requesting process.

In this work, we present a hybrid-parallel particle advection system, coupled with a local load balancing method based on diffusive load balancing \cite{Boillat1990,Cybenko1989}. The processes load the blocks of their first neighbors in addition to their own. This partial redundancy enables efficient application of nearest-neighbor family of diffusive load balancing algorithms based on the framework defined by Cybenko \cite{Cybenko1989}. We introduce two scheduling methods, considering the requirements of the specific problem of particle advection. 


\section{Related Work}
\label{section/related_work}
The approach presented in \cite{Pugmire2009} is an earlier application of hybrid parallelism to particle advection. The processes are divided into groups, each containing a master load balancing process which dynamically distributes the data and the particle subsets to the rest. The method overcomes the scaling issues of a central load balancer, by introducing multiple centers. 

The recent approach presented in \cite{Zhang2018a} dynamically adjusts the blocks assigned to the processes at each round of advection. The method constructs a dynamic access dependency graph (ADG) from the number of transmissions between neighboring processes. This information is combined with the particle sets to resize and relocate the blocks assigned to the processes. 

The dynamic K-D tree decomposition presented in \cite{Zhang2018b} is an example of a hybrid approach, combining static data parallelism with dynamic task parallelism. In this method, each process statically loads a larger data block than the region it is assigned to. At each advection round, a constrained K-D tree is constructed based on the current particle set, whose leaves are used to resize the regions of the processes, up to the boundary of their data blocks.

The approach presented in \cite{Binyahib2019a} adapts the lifeline-based load balancing method to the context of parallel particle advection. The method builds on work requesting \cite{Mueller2013}, in which processes with low load ask processes with higher load to share (often half of) their load. In contrast to random selection of the process to request work from, this approach first constructs a lifeline graph of processes. The requests are not random, and made to the adjacent nodes in the graph, which may forward the request recursively to their neighbors.  


\section{Method}
\label{section/method}
In this section, we first provide an overview of baseline hybrid-parallel particle advection and a review of the mathematical framework for diffusive load balancing as defined in \cite{Cybenko1989}. We then introduce two methods that are based on diffusive load balancing. 

\subsection{Hybrid-Parallel Particle Advection}
\label{section/method/hybrid_parallel_particle_advection}
The approach to achieve hybrid parallelism is to parallelize over data among the processes, and parallelize over the tasks among the threads of each process. Within the context of particle advection, this corresponds to distribution of the vector field blocks as well as the related particle subsets to the processes, and integrating the latter using an array of threads. 

The following pseudocode outlines the kernel ran by each process. We also provide two points of entry for modular implementation of the load balancing algorithms which will be discussed in the next section.

\begin{algorithm}
\label{algorithm/1}
\SetKwInOut{Input}{Input}
\SetKwInOut{Output}{Output}
\ResetInOut{Output}
\underline{function Advect} $(p, v)$\;
\Input{Particles $p$ and vector fields $v$}
\Output{Advected particles $p'$ and integral curves $i$}
\While{!$check\_completion$(p)}{
  $load\_balance\_distribute$(p);\\
  r = $compute\_round\_info$(p, i);\\
  $allocate\_integral\_curves$(p, i, r);\\
  $integrate$(v, p, p', i, r);\\
  $load\_balance\_collect$(r);\\
  $out\_of\_bounds\_distribute$(p, r);
}
\caption{The advection kernel.}
\end{algorithm}

The outermost condition, \textit{check\_completion}, consists of two global collective operations, a gather for retrieving the current active particle counts of the processes, and a broadcast that consists of a single Boolean evaluating to true only when all gathered particle counts are zero. 

The \textit{load\_balance\_distribute} is a function that provides the active particles (as well as the index of the process), and expects the implementer to freely call global or neighborhood collective operations to balance the load of the process for the upcoming round.

The function \textit{compute\_round\_info} generates the metadata to conduct an advection round, setting the range of particles, tracking vertex offsets and strides for the integral curves, as well as creating a mapping of out-of-bound particles to the neighboring processes.

The \textit{allocate\_integral\_curves} (re-)allocates the vertex array of the integral curves, expanding it by the number of particles to trace this round times the longest iteration among those particles. Despite potentially wasteful usage of memory, one allocation per round is significantly more efficient than using resizable nested vectors of points. Besides, it is possible to prune the curves based on a reserved vertex either at the end of each round or after the advection kernel has finished, which may also be done in parallel using the execution policies recently introduced to the standard template library.

The \textit{integrate} iterates through the particles of the round, sampling the corresponding vector field at the position of the particle, and integrates the curve based on the interpolated vector using a scheme such as the Runge-Kutta family of methods. 

The \textit{load\_balance\_collect} is an optional function providing a way to recall the particles that have been transmitted to other processes during load balancing. In the case of diffusive load balancing, particles that have gone out-of-bounds after being load balanced to a neighboring process have to be returned to the original process prior to the out-of-bounds distribution stage. This is due to the fact that the neighboring process does not have any information on or topological connection to the neighbors of the original process except itself.

The \textit{out\_of\_bounds\_distribute} is the final stage, in which particles reaching the boundary of a neighbor during the round are transmitted to the corresponding neighboring processes using local collectives.


\subsection{Load Balancing}
\label{section/method/load_balancing}
A set of processes arranged in a Cartesian grid may be interpreted as a connected graph where the nodes are the processes and the edges are the topological connections between the processes. Assume the processes are labelled from 0 to $n$. Define $w^t$ as an n-vector quantifying the work distribution, so that $w_i^t$ is the amount of work to be done by process $i$ at time $t$. The diffusion model for dynamic load balancing as described by Cybenko in \cite{Cybenko1989} has the following form:
\begin{equation}
  w_i^{t+1} = w_i^t + \sum_j \alpha_{ij} (w_j^t - w_i^t) + \eta_i^{t+1} - c
\end{equation}
where $\alpha_{ij}$ are coupled scalars, evaluating to non-zero only when process $i$ and $j$ are topologically connected. Within a Cartesian grid, this corresponds to the immediate neighbors excluding the diagonals. The sum implies that the processes $i$ and $j$ compare their workloads at time $t$ and transmit $\alpha_{ij} (w_j^t - w_i^t)$ pieces of work among each other. The work is transmitted from $j$ to $i$ if the quantity is positive, and vice versa if negative. The term $\eta_i^{t+1}$ corresponds to new work generated at process $i$ at time $t$. The term $c$ describes the amount of work performed by the process between time $t$ and $t+1$. 

In the next section, we propose a method to compute $\alpha_{ij}$ dynamically based on several local averaging operations, guaranteeing an improved distribution of the workload. A central requirement for the following algorithms for computation of $\alpha_{ij}$ is the availability of information from the first neighbors. Each process is required to additionally have access to the data blocks of its neighbors. We furthermore communicate the local workload for the next round of particle advection to all neighbors dynamically at the beginning of each load balancing step. 


\begin{algorithm}
\label{algorithm/2}
\SetKwInOut{Input}{Input}
\SetKwInOut{Output}{Output}
\ResetInOut{Output}
\underline{function LMA} $(load, neighbor\_loads)$\;
\Input{Local workload amount $load$, and vector of neighbors' workload amounts $neighbor\_loads$.}
\Output{Vector of workload amounts to be sent to each neighbor $outgoing\_loads$.}
$contributors$ = vector<bool>($neighbor\_loads$.size());\\
$mean$  = $load$;\\
\Do{There are neighbors above $mean$ in $contributors$}
{
  $contributors$.fill(false);\\
  $sum$ = $load$;\\
  $count$ = 1;  \\
  \For{i=0; i < $neighbor\_loads$.size(); i++}
  {
    \If{$neighbor\_loads[i]$ < $mean$}
    {
      $contributors[i]$ = true;\\
      $sum$ += $neighbor\_loads[i]$;\\
      $count$++;
    }
  }
  $mean$  = $sum$ / $count$;\\
} 
$ $ \\
$outgoing\_loads$ = vector<uint>($neighbor\_loads$.size());\\
$outgoing\_loads$.fill(0);\\
\For{i=0; i < $neighbor\_loads$.size(); i++}
{
  \If{$contributors[i]$ == true}
  {
    $outgoing\_loads[i]$ = $mean$ - $neighbor\_loads[i]$;
  }
}
\Return{$outgoing\_loads$};
\caption{The lesser mean assignment.}
\end{algorithm}

\begin{figure}[t]
  \centering
  \includegraphics[width=\linewidth]{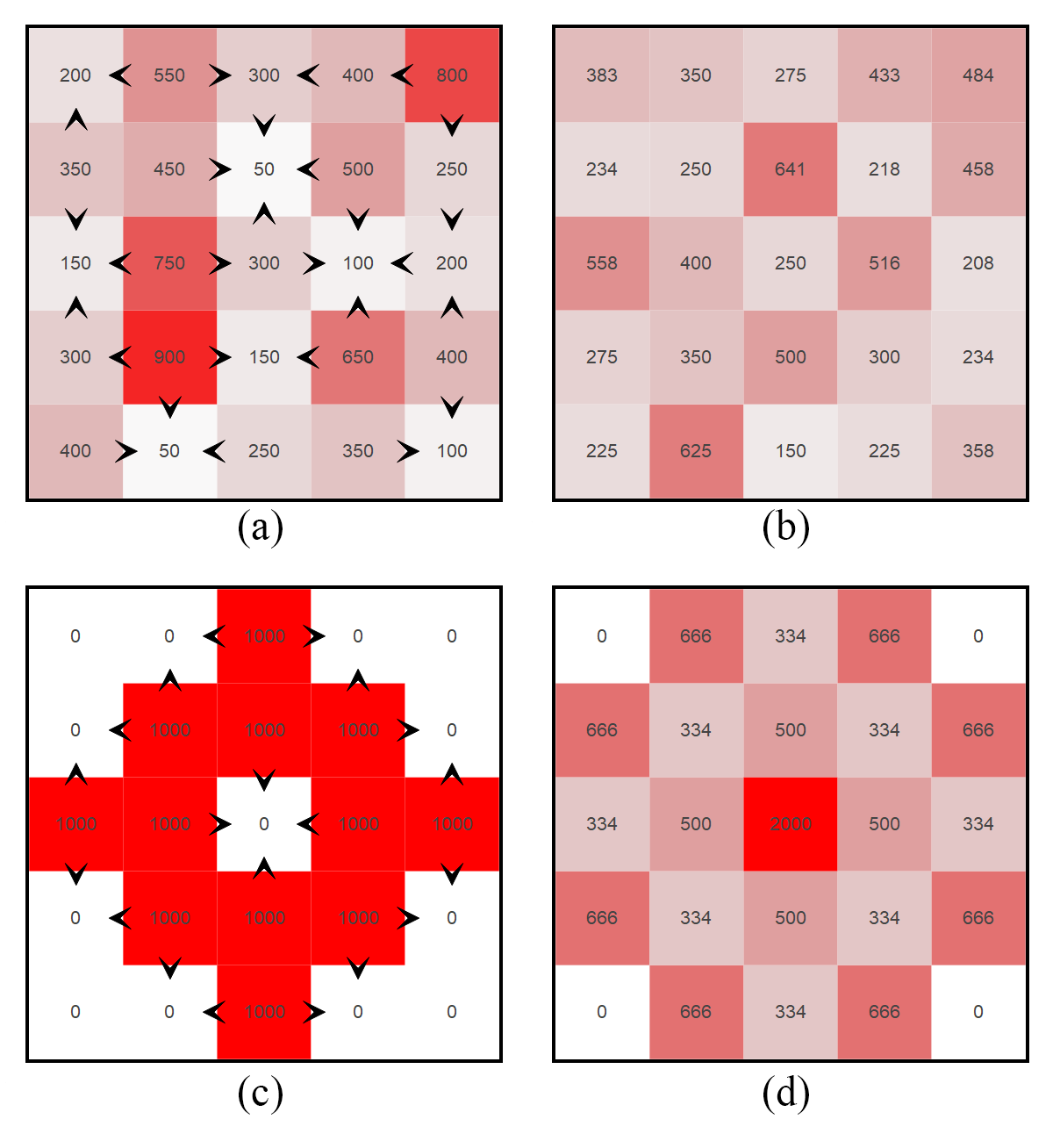}
  \caption{An illustration of particles (loads) assigned to a grid of processes. The lesser mean assignment balances the state (a) to state (b). Certain cases in which a lesser loaded process is surrounded by greater loaded processes such as (c) may lead to over-balancing as seen in (d).}
  \label{figure/2.png}
\end{figure}

\subsubsection{Lesser Mean Assignment}
\label{section/method/load_balancing/lma}

Lesser mean assignment (LMA), detailed in Algorithm \ref{algorithm/2} and illustrated in Figure \ref{figure/2.png}, attempts to equalize the workload of the local process and its lesser loaded neighbors. Upon the transmission of neighbor workloads, the mean of the local process and any neighbor with less load is computed. The mean is revised iteratively, removing any neighbors contributing to it which have more workload than it. Finally, we transmit the difference of each contributing neighbor from the mean to the corresponding processes.

The method could be interpreted as a unidirectional flow of workload from more to less loaded processes. In contrast to the original load balancing approach by Cybenko \cite{Cybenko1989}, which fixes the diffusion parameter to $1 - \frac{2}{dimensions+1}$ of the difference from each neighboring process, the method scans the whole neighborhood for an optimal local distribution.

Due to unconditional flow of load from greater loaded neighbors, the lesser loaded processes potentially suffer from an issue of \textit{over-balancing}, in which they receive a large sum of particles when surrounded by greater loaded neighbors. Since second-neighbor information is not available, the greater loaded processes are unable to take the neighborhood of the lesser loaded processes into account. The issue potentially results in cases in which the global maximum load increases, as seen in the second row of Figure \ref{figure/2.png}. 

\begin{algorithm}
\label{algorithm/3}
\SetKwInOut{Input}{Input}
\SetKwInOut{Output}{Output}
\ResetInOut{Output}
\underline{function GLLMA} $(load, neighbor\_loads)$\;
\Input{Local workload amount $load$, and vector of neighbors' workload amounts $neighbor\_loads$.}
\Output{Vector of workload amounts to be sent to each neighbor $outgoing\_loads$.}
$contributors$ = vector<bool>($neighbor\_loads$.size());\\
$mean$  = $load$;\\
\Do{There are neighbors below $mean$ in $contributors$}
{
  $contributors$.fill(false);\\
  $sum$ = $load$;\\
  $count$ = 1;  \\
  \For{i=0; i < $neighbor\_loads$.size(); i++}
  {
    \If{$neighbor\_loads[i]$ > $mean$}
    {
      $contributors[i]$ = true;\\
      $sum$ += $neighbor\_loads[i]$;\\
      $count$++;
    }
  }
  $mean$  = $sum$ / $count$;\\
} 
$ $ \\
$total\_quota$ = $mean$ - $load$;\\
$quotas$ = vector<uint>($neighbor\_loads$.size());\\
$quotas$.fill(0);\\
\For{i=0; i < $neighbor\_loads$.size(); i++}
{
  \If{$contributors[i]$ == true}
  {
    $quotas[i]$ = $total\_quota \cdot neighbor\_loads[i] / (sum - load)$;
  }
}
neighborhood\_all\_gather($quotas$);\\
$ $ \\
$outgoing\_loads = LMA(load, neighbor\_loads);$\\
\Return{pairwise\_minimum($outgoing\_loads$, $quotas$)};
\caption{The greater limited lesser mean assignment.}
\end{algorithm}

\begin{figure}[t]
  \centering
  \includegraphics[width=\linewidth]{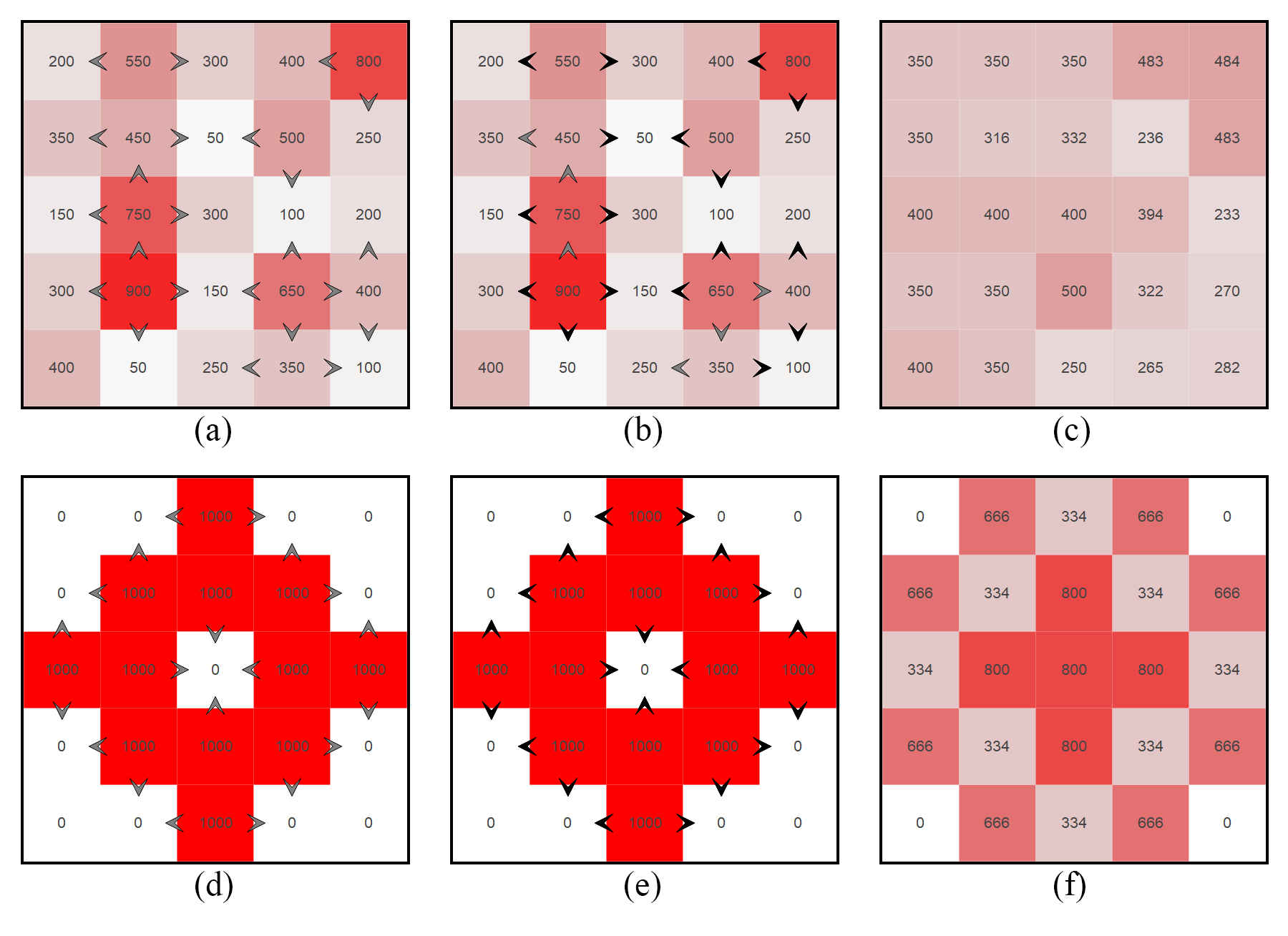}
  \caption{The greater-limited lesser mean assignment first declares quotas for the greater loaded neighbors, limiting the flow as in (a, d), followed by lesser mean assignment, balancing from (b, e) to (c, f). Cases that lead to over-balancing are averted by the initial quota process as seen in the second line, still providing a 20\% boost to round run-time.}
  \label{figure/3.png}
\end{figure}

\subsubsection{Greater-Limited Lesser Mean Assignment}
\label{section/method/load_balancing/gllma}

Greater-limited lesser mean assignment (GL-LMA) is detailed in Algorithm \ref{algorithm/3} and illustrated in Figure \ref{figure/3.png}. It introduces a preprocessing step to LMA, utilizing the higher loaded neighbor information in an effort to prevent the over-balancing effect.

The processes start by computing the mean of their own load and any neighbor with \textit{greater} load. The mean is revised iteratively, omitting neighbors with lesser load, ensuring that the local process is the only one with less load than it. The processes then define their total quota, that is the total number of particles they would accept from their neighbors, as the difference between the greater mean and their local load. Each neighbor is assigned a part of this quota, proportional to their contribution to the greater mean.

The quotas are transmitted to the associated neighbors, through a local collective operation that we factor in as \textit{neighborhood\_all\_gather} in Algorithm \ref{algorithm/3}. This function may be realized using \textit{MPI\_Neighbor\_allgather} or implemented manually as a series of local communications if MPI 3.0 is not available. Upon the transmission of the quotas, the processes apply a round of LMA, but limit the amount of outgoing particles by the quotas received from the neighbors.

The approach requires a third local collective operation in addition to the load and particle transmission stages, yet effectively prevents the \textit{over-balancing} effect LMA is prone to. In an overview, each process first sets quotas for its greater loaded neighbors, and then sends particles to its lesser loaded neighbors. The approach indirectly makes second neighbor information available to the processes, since they are now partially aware of the neighborhood of their lesser loaded neighbors through the quota received from them.

\section{Performance}
\label{section/performance}
This section details the experiments we have conducted to measure the runtime performance and scaling of the presented load balancing algorithms. We first enumerate and describe the variables of the particle advection pipeline in Section \ref{section/performance/variables}, and then construct a series of experiments by systematically varying them in Section \ref{section/performance/experiments}. The metrics accompanying the measurements are detailed in Section \ref{section/performance/metrics}.

\subsection{Variables}
\label{section/performance/variables}

\subsubsection{Number of Processes}
The number of processes is an axis of measurement, and is defined in terms of compute nodes rather than cores in this work. The process count also controls the number of partitions in a data-parallel sense; each process is responsible for one partition in the baseline approach, and for seven partitions (in 3D) in the diffusive load balancing approaches. Increasing the number of nodes refines the domain partitioning, which accelerates the compute.

\subsubsection{Load Balancing Algorithm}
A central variable is the load balancing algorithm. The pipeline supports four methods which are no diffusion, constant diffusion, lesser mean assignment and greater-limited lesser mean assignment. Each experiment is performed on all four methods regardless of measurement context, in order to quantify the benefits and limitations of the presented algorithms with respect to the baseline.
\begin{figure*}[t]
  \centering
  \includegraphics[width=\linewidth]{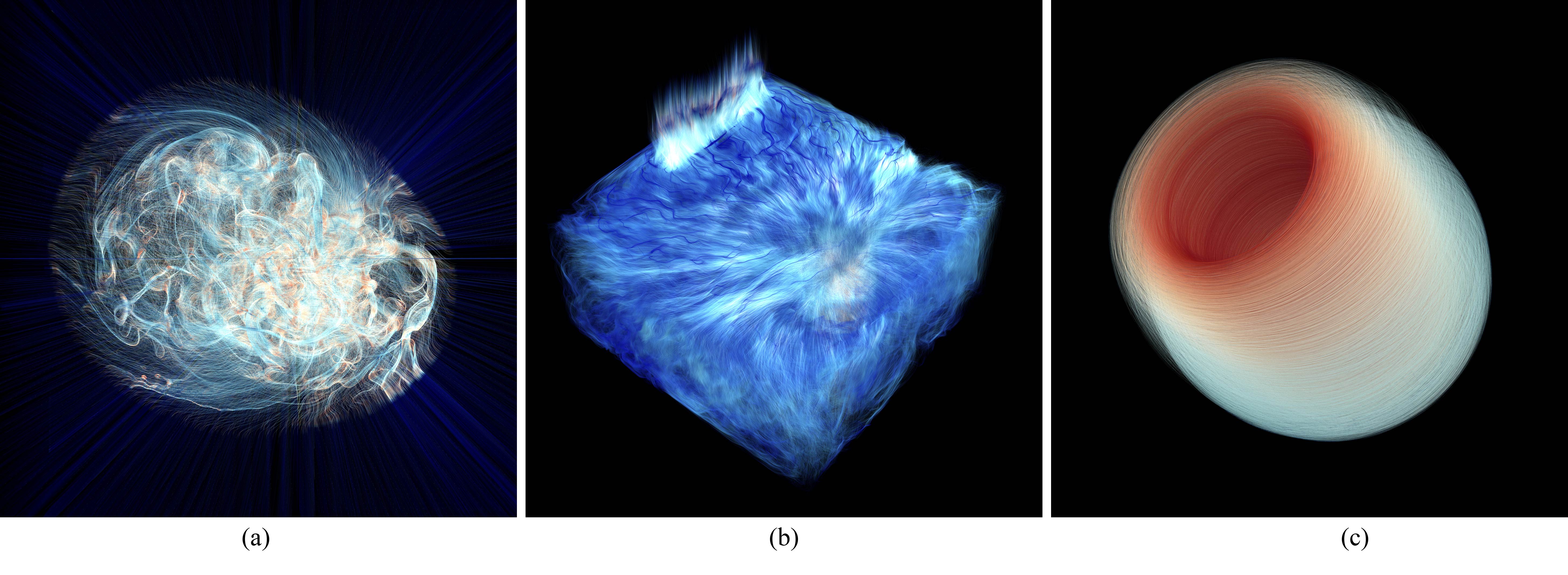}
  \caption{The astrophysics dataset (a) is a simulated magnetic field around a core-collapse supernova, with complex trajectories concentrated near the center. The thermal hydraulics dataset (b) is a simulation featuring two sources pumping water at different temperatures into a rigid box. The nuclear fusion dataset (c) is a simulation of the magnetic field within a Tokamak device, consisting of many orbital trajectories. All datasets comprise several GBs, streamlines are extracted by our approach and ray traced via \cite{Wald2017}. Color indicates vector magnitude.}
  \label{figure/4.jpg}
\end{figure*}

\subsubsection{Dataset}
The dataset is a variable since the complexity of the input vector field impacts the performance of particle advection \cite{Pugmire2009}. We have obtained three datasets covering the domains of astrophysics, thermal hydraulics and nuclear fusion from the Research Group on Computing and Data Understanding at eXtreme Scale at the University of Oregon. Each dataset contains features which lead to distinct particle behavior as seen in Figure \ref{figure/4.jpg}. The datasets are identical to the ones presented in \cite{Binyahib2019a}, which we refer the reader to for further detail.

\subsubsection{Dataset Size}
The size of the dataset is a separate variable. It is in a linear relationship with the size of the partitions and hence influences performance. We scale each dataset to $1024^3$, $1536^3$, $2048^3$ voxels, leading to $12.8$GB, $43.4$GB and $103$GB floating-point triplets respectively.

\subsubsection{Seed Set Distribution}
The seed set distribution is another variable, as it has a direct impact on load imbalance. The particles are uniformly generated within an axis-aligned bounding box (AABB) located at the center of the dataset. Scaling the AABB enables concentrating the particles towards the center and vice versa. We limit the distributions to uniform scaling by $0.25$, $0.5$ and $1.0$ (the complete dataset).

\subsubsection{Seed Set Size}
The seed set size is the final variable, which corresponds to the amount of particles/work. It is controlled through \textit{stride}, a vector parameter describing the distance between adjacent particles in each dimension. A stride of $[1, 1, 1]$ implies one particle per voxel. We limit the strides to $[8, 8, 8]$, $[8, 8, 4]$, $[8, 4, 4]$, $[4, 4, 4]$, a sequence which may be interpreted as cumulatively doubling the amount of work across Z, Y and X. 

The rest of the variables are fixed for all experiments: The integrator is set to Runge-Kutta 4, with a constant step size of $0.001$. Maximum iterations per particle is set to $1000$. Note that these settings are independent of the dataset size as long as the domain is scaled to the $[0, 1]$ range in each dimension. Despite being adjustable, particles-per-round are set to $10$ million in order to reduce benchmark combinations.

\subsection{Experiments}
\label{section/performance/experiments}
We assess the strong and weak scaling of the system, applying each algorithm to each dataset. Furthermore, we record and present the per-round load imbalance of the algorithms in a fixed setting. We finally scan the parameter space, assessing the impact of each variable in isolation.

\begin{figure*}
  \centering
  \includegraphics[width=\linewidth]{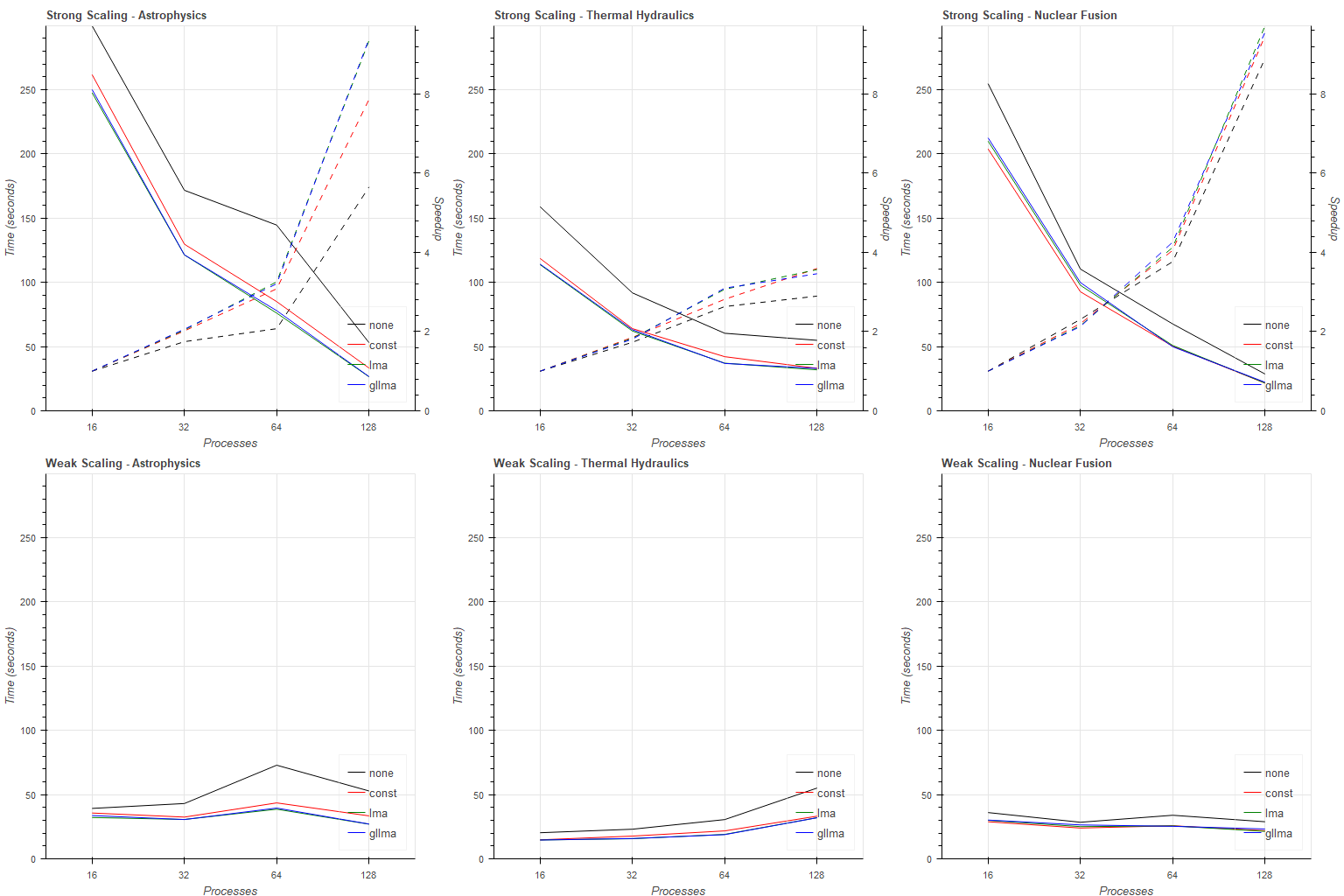}
  \caption{Strong (top) and weak (bottom) scaling benchmarks. Black lines are no diffusion, red lines are constant diffusion, green lines are LMA, blue lines are GL-LMA. Datasets vary from left to right; astrophysics, thermal hydraulics, nuclear fusion. Solid lines are total advection times, dashed lines are speedups.}
  \label{figure/5.png}
\end{figure*}

\subsubsection{Strong Scaling}
\label{section/performance/experiments/strong_scaling}
The number of processes is varied among 16, 32, 64, 128. Every other variable is fixed: The dataset sizes are set to $1024^3$, the seed set distribution spans the whole domain (uniform scaling of the AABB by $1.0$) and the stride is $[4, 4, 4]$, leading to $16.78$ million particles. The experiment is performed independently for each algorithm and dataset, leading to $nodes \cdot algorithms \cdot datasets = 4 \cdot 4 \cdot 3 = 48$ measurements. We record the total duration of particle advection for each process. 

\subsubsection{Weak Scaling}
\label{section/performance/experiments/weak_scaling}
The number of processes is varied among 16, 32, 64, 128 simultaneously with stride, which is varied among $[8, 8, 8]$, $[8, 8, 4]$, $[8, 4, 4]$, $[4, 4, 4]$. Both the number of processes and the amount of work is doubled, as the stride leads to $2^{21}$, $2^{22}$, $2^{23}$ and $2^{24}$ particles respectively. The rest of the variables are fixed, identical to the strong scaling benchmarks: The dataset sizes are $1024^3$ and the particle distribution spans the whole dataset. The experiment is performed independently for each algorithm and dataset, leading to $nodes \cdot algorithms \cdot datasets = 4 \cdot 4 \cdot 3 = 48$ additional measurements. We record the total duration of particle advection for each process.

\subsubsection{Load Balancing}
\label{section/performance/experiments/load_balancing}
The number of processes is fixed to 16. The dataset sizes are set to $1024^3$. The seed set distribution contains half of the domain in each dimension (uniform scaling of the AABB by $0.5$) and the stride is set to $[4, 4, 4]$, yielding $16.78$ million particles. The particles are concentrated near the center of the dataset, leading to significant load imbalance from the start. The experiment is performed independently for each algorithm and dataset, leading to $algorithms \cdot datasets = 4 \cdot 3 = 12$ measurements. We record the duration of each stage of each round per-process. 

\subsubsection{Parameter Space}
\label{section/performance/experiments/parameter_space}
The number of processes is varied among 16, 32, 64, 128 in addition to one of: dataset complexity, dataset size, seed distribution, seed stride. Each of the latter variables have at least three settings described in Section \ref{section/performance/variables}, which are tested in isolation. Note that when varying dataset size, we proportionally vary stride in order to keep the particle count constant. When fixed, the dataset is set to astrophysics, the dataset size is set to $1024^3$, the seed distribution is set to $1.0$ and the seed stride is set to $[4, 4, 4]$. The experiment is performed for each algorithm, leading to $nodes \cdot algorithms \cdot (variables \cdot variable\_options) = 4 \cdot 4 \cdot (4 \cdot 3) = 192$ measurements. We record the total duration of particle advection for each process. Note that the dataset complexity configurations are equivalent to the strong scaling tests, yet are recorded separately and included for comparison to other parameters. 

\subsection{Metrics}
\label{section/performance/metrics}
For each scaling measurement we compute the speedup from the time measurements. The speedup is always defined in terms of nodes, rather than cores. It is furthermore relative to the first measurement, since the problem sizes often exceed the capabilities of the serial application. That is, for the strong scaling benchmarks ran on $N_i<...<N_j$ nodes, speedup is defined as: 
\begin{equation*}
  S(N) = \frac{T(N_i)}{T(N)}
\end{equation*}
where $T(N)$ is the duration of the application with $N$ nodes.

For the load balancing measurements, we compute the load imbalance factor in each round as:
\begin{equation*}
  LIF = \frac{load_{max}}{load_{avg}}
\end{equation*}
which is equal to $1$ when all processes have an equal amount of work. The metric denotes the distance of the current load distribution from the optimal state where each process has exactly average workload.

\section{Results \& Discussion}
\label{section/results_and_discussion}
This section presents the results accompanied by a comparative discussion on the performance of the four algorithms under varying conditions.

The performance benchmarks are ran on the RWTH Aachen CLAIX-2018 (c18m) compute cluster, which offers up to 1024 nodes containing 2x24 Intel Skylake Platinum 8160 cores, along with 192GB of memory per node. 

The tests are conducted using 16 to 128 nodes. The nodes are exclusively reserved for the application, in order to eliminate any effects due to resource consumption of other processes. Each node is configured to run one instance of the MPI application, which saturates all 48 cores and the complete memory. The number of cores and memory per node are fixed throughout the tests.

\begin{figure*}[!t]
  \centering
  \includegraphics[width=\linewidth]{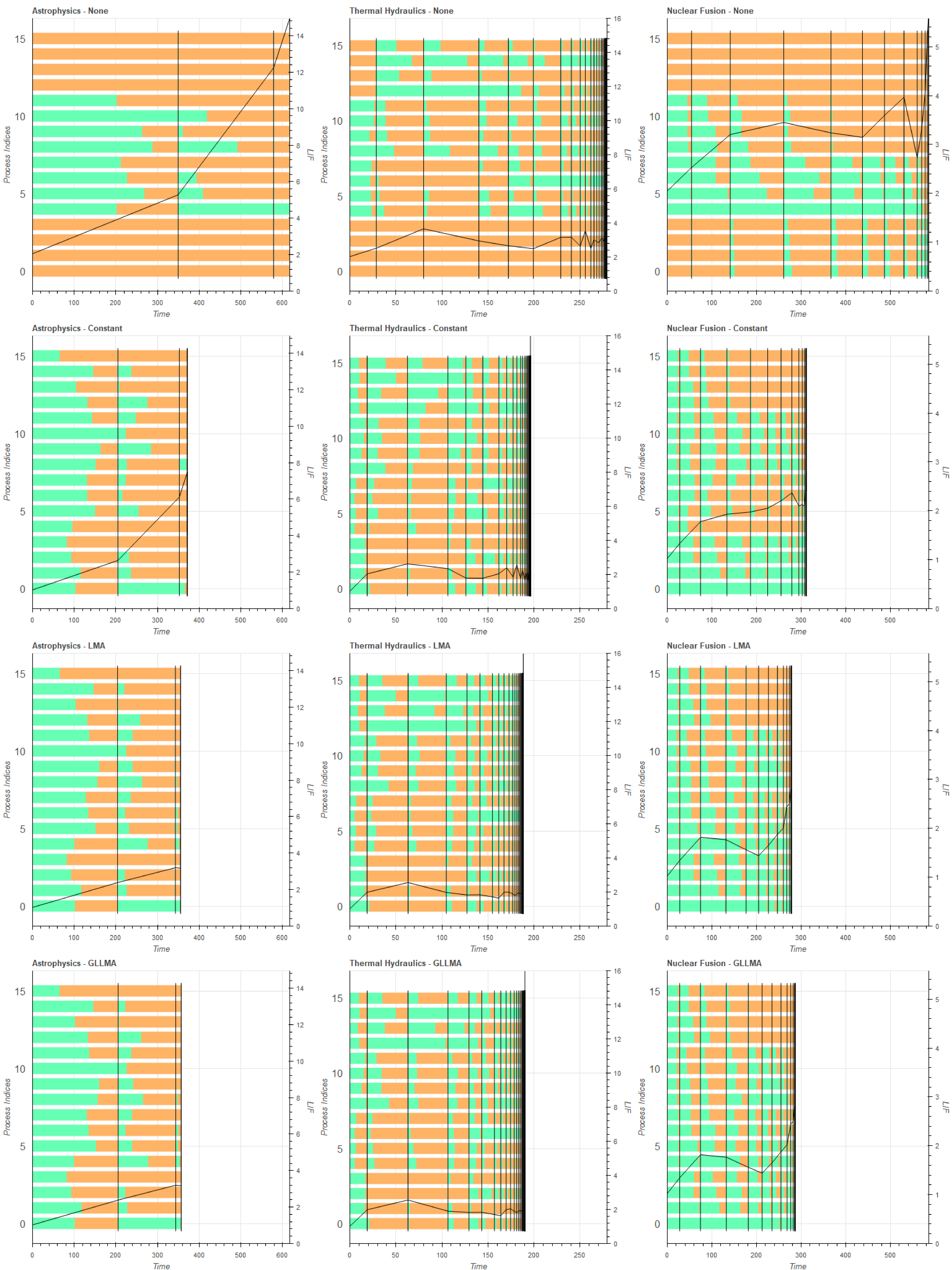}
  \caption{Gantt charts of round times per process. Green is active compute, orange is idle. Datasets vary from left to right; astrophysics, thermal hydraulics, nuclear fusion. Load balancing algorithms vary from top to bottom; no diffusion, constant diffusion, LMA, GL-LMA. Load imbalance factors are overlaid as line plots.}
  \label{figure/6.jpg}
\end{figure*}

\begin{figure*}[!t]
  \centering
  \includegraphics[width=\linewidth]{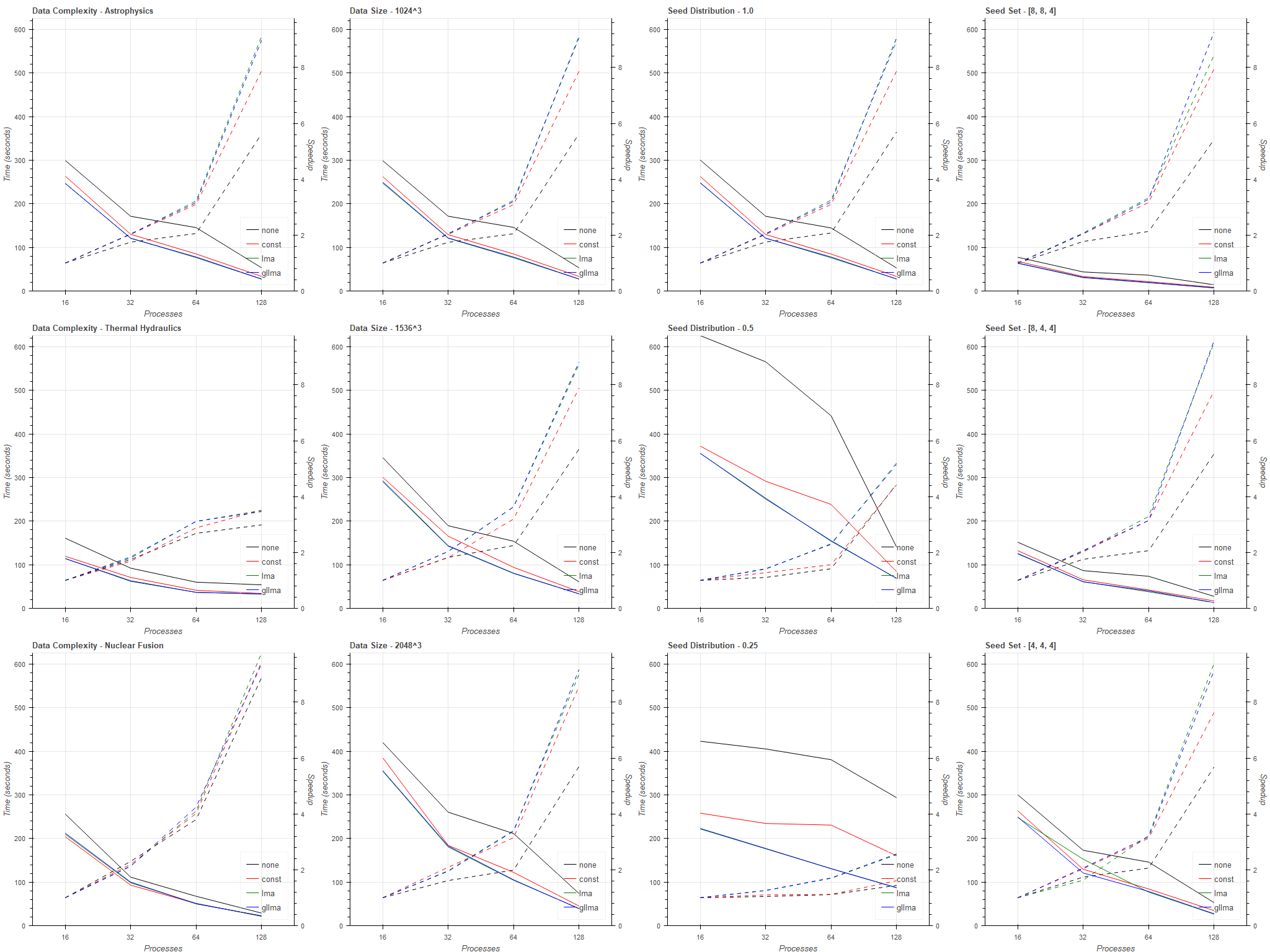}
  \caption{Parameter space benchmarks. From left to right; dataset type, dataset size, seed distribution, seed set size. Black lines are no diffusion, red lines are constant diffusion, green lines are LMA, blue lines are GL-LMA. Solid lines are total advection times, dashed lines are speedups.}
  \label{figure/7.png}
\end{figure*}

\subsection{Strong Scaling}
\label{section/results/strong_scaling}
The strong scaling benchmarks are presented as line plots mapping node counts to total advection times in the first row of Figure \ref{figure/5.png}. The plots show asymptotic behavior, with the diffusive approaches consistently outperforming the baseline. The LMA and GL-LMA display nearly identical performance implying that overbalancing does not occur. Both approaches outperform constant diffusion in 10 out of 12 cases. The exceptions are the 16 and 32 node runs on the nuclear fusion dataset, which we attribute to the coarseness of domain partitioning at low node counts. The total advection times converge with increasing node counts. 

The baseline yields a relative speedup of $2.07$ for the 64 node run on the astrophysics dataset, significantly lower than the diffusive approaches which provide a minimum of $3.11$. Irregularities regarding this configuration also appear in the latter tests, which may indicate a relationship to the domain partitioning for the given node count.

\subsection{Weak Scaling}
\label{section/results/weak_scaling}
The weak scaling benchmarks map node count - stride pairs to total advection times as presented in the second row of Figure \ref{figure/5.png}. The astrophysics set leads to non-constant behavior with the 64 node configuration, although this effect is largely suppressed by the diffusive load balancing approaches.

The thermal hydraulics dataset displays an increase in total runtime with increasing nodes and work, becoming apparent in the 128 node configuration. The dataset penalties static domain partitioning, as it contains long trajectories covering the whole domain. Refining the domain subdivision leads to increased communication across nodes, which negatively affect runtime, a pattern which is also seen in Figure \ref{figure/6.jpg}. The nuclear fusion dataset displays constant behavior.

\subsection{Load Balancing}
\label{section/results/load_balancing}
The load balancing benchmarks are presented as Gantt charts mapping per-stage advection times to the nodes in Figure \ref{figure/6.jpg}. Note that due to the seed distribution being set to the center $0.5$ of the dataset, the outer processes are expected to initially idle. The GL-LMA nearly halves the total time advection in the astrophysics dataset, bringing $617$ seconds down to $356$. Similar results are observed for the thermal hydraulics and nuclear fusion datasets.

The LMA outperforms GL-LMA in cases where many short rounds are involved. The per-round overhead of the latter, along with the limits it applies, leads to longer total advection times.

\subsection{Parameter Space}
\label{section/results/parameter_space}
The parameter space experiments are presented as line plots in Figure \ref{figure/7.png} and are in agreement with the strong scaling benchmarks. The LMA and GL-LMA achieve the lowest advection time in 46 of the 48 measurements, along with the highest speedups for all 16 configurations.

The astrophysics and nuclear fusion datasets appear to greatly benefit from increased process counts, yielding relative speedups of $8.9$ and $9.37$ for the 128 node setting whereas this is limited to $3.45$ for thermal hydraulics. The dataset contains empty regions above and below the box, which leads to uneven partitioning in the associated axis.

Changes to dataset and seed set size display a constant response to total advection time, yet seed distribution appears to have a great effect as seen in the third column. LMA and GL-LMA are more resilient to increase in locality, compared to the baseline and constant case. Overbalancing occurs in the 32 node setting of the densest seed set distribution, yet is averted by GL-LMA as seen in the fourth column.

\section{Conclusion} 
\label{section/conclusion_and_future_work}
We have presented a distributed particle advection system along with a local load balancing method based on partial data replication among process neighborhoods. Two local scheduling methods based on diffusive load balancing have been applied to the problem of parallel particle advection. The performance has been assessed through strong and weak scaling benchmarks as well as load imbalance metrics. The results are shown to improve total runtime and are promising regarding scaling, since the method exclusively operates on local neighborhoods of processes, avoiding any global communication. 


\section{Acknowledgements}
\label{section/acknowledgements}
The authors gratefully acknowledge the computing time granted by the NHR4CES Resource Allocation Board and provided on the supercomputer CLAIX at RWTH Aachen University as part of the NHR4CES infrastructure. The calculations for this research were conducted with computing resources under the project rwth0432.

\bibliographystyle{alpha}
\bibliography{bibliography}

\end{document}